\documentclass{article}
\usepackage{amsmath}

\input{tcilatex}

\begin{document}

\title{Derivation of the spatio-temporal model equations for the
thermoacoustic resonator.}
\author{Isabel P\'{e}rez-Arjona$^{1}$, V\'{\i}ctor J. S\'{a}nchez-Morcillo$%
^{1}$ \and and Germ\'{a}n de Valc\'{a}rcel$^{2}$. \\
$^{1}$I{\small nstitut de Investigaci\'{o} per a la Gesti\'{o} Integrada de
las Zones Costaneres (IGIC)}\\
{\small \ Universitat Polit\`{e}cnica de Val\`{e}ncia}\\
{\small Carretera Nazaret-Oliva s/n}\\
{\small 46730 Grau de Gandia (Val\`{e}ncia) SPAIN}\\
$^{2}${\small Departament d'\`{O}ptica- Universitat de Val\`{e}ncia}}
\maketitle

\begin{abstract}
We derive the model equations describing the thermoacoustic resonator, that
is, an acoustical resonator containing a viscous medium inside. Previous
studies on this system have addressed this sytem in the frame of the
plane-wave approximation, we extend the previous model to by considering
spatial effects in a large aperture resonator. This model exhibits pattern
formation and localized structures scenario.
\end{abstract}

\section{Derivation of the model}

The physical system we consider is an acoustic resonator composed of two
parallel walls containing a viscous fluid medium inside (e.g. glycerine).
The case of sound beam propagating in viscous media is characterized by a
strong absorption, being the sound velocity dependent on the fluid
temperature, which results in an addicional nonlinearity mechanism from
thermal origin.

\subsection{Sound propagation in a viscous medium}

The propagation of sound in a viscous, heat-conducting medium is described
by the equations of mechanics of continuum media \cite{Naugolnykh}, $i.e.$
the continuity equation%
\begin{equation}
\frac{\partial \rho }{\partial t}+\rho _{0}\mathbf{\nabla }\cdot \mathbf{v}%
=0,  \label{continuityEq}
\end{equation}%
the momentum transfer equation%
\begin{equation}
\frac{\partial \mathbf{v}}{\partial t}+\frac{1}{\rho _{0}}\mathbf{\nabla }%
p=\rho _{0}\mu \nabla ^{2}\mathbf{v}+\left( \mu _{B}+\frac{\mu }{3}\right) 
\mathbf{\nabla }\left( \mathbf{\nabla }\cdot \mathbf{v}\right) ,
\label{momentumtrans}
\end{equation}%
and the heat transport equation%
\begin{equation}
\rho _{0}c_{p}\frac{\partial T}{\partial t}-\kappa \nabla ^{2}T=\mu \frac{%
\partial v_{i}}{\partial x_{k}}\left( \frac{\partial v_{i}}{\partial x_{k}}+%
\frac{\partial v_{k}}{\partial x_{i}}-\frac{2}{3}\delta _{ik}\frac{\partial
v_{l}}{\partial x_{l}}\right) +\mu _{B}\left( \mathbf{\nabla }\cdot \mathbf{v%
}\right) ^{2},  \label{heattransport}
\end{equation}%
where $\rho $ and$~p$ represent, respectively, the density and acoustic
pressure deviations from the equilibrium state (denoted by $\rho _{0}$ and $%
p_{0})$, $\mathbf{v}$ is the fluid particle velocity. The constants $\mu ,$ $%
\mu _{B},\kappa $ and $c_{p}$ are the coefficients of shear and bulk
viscosity, the thermal conductivity and the specific heat at constant
pressure of the fluid, respectively. In Eq.(\ref{heattransport}), $T$
denotes the perturbation of the medium temperature due to the acoustic wave,
and $x_{i}$ the spatial coordinates.

Equations (\ref{continuityEq})-(\ref{heattransport}) must be complemented
with the equation of state relating the medium density with the other
variables. For a linear medium (small amplitudes) it obeys the simple form $%
p=c_{0}^{2}\rho .$ In a highly absorbing medium, the thermal effects are
dominant and the hydrodynamic nonlinearity can be neglected. Furthermore,
the speed of sound depends on the average value (over one period) of the
temperature $T_{av}=T_{0}+\bar{T},$ where $T_{0}$ is the temperature at
equilibrium, so the equation of state takes the form $\rho =p/c_{0}^{2}(\bar{%
T}).$ Expanding at leading order it follows%
\begin{equation}
\rho =\frac{1}{c_{0}^{2}}p-\frac{\gamma _{p}}{c_{0}^{2}}p\bar{T}.
\label{stateEq}
\end{equation}%
where $\gamma _{p}=(2/c_{0})\left( \frac{\partial c}{\partial T}\right)
_{T=T_{eq}}$

We assume that the fields can be described by quasi-plane waves, i.e. they
take the form 
\begin{subequations}
\label{quasiplane}
\begin{eqnarray}
p &=&\frac{1}{2}P(\mathbf{r},t)e^{i\omega t}+c.c., \\
\mathbf{v} &=&\frac{1}{2}\mathbf{u}(\mathbf{r},t)e^{i\omega t}+c.c., \\
T &=&\frac{1}{2}T(\mathbf{r},t)e^{i\omega t}+c.c.+\bar{T}(\mathbf{r},t),
\end{eqnarray}%
where a slowly changing temperature $\bar{T}$, responsible for the thermal
self-action of the waves on $v=v_{z}$, is added. Taking into account the
slowly varying envelope condition 
\end{subequations}
\begin{equation}
\frac{\partial X_{i}}{\partial x_{j}}<<kX_{i},~\frac{\partial X_{i}}{%
\partial t}<<\omega X_{i},  \label{SVEA}
\end{equation}%
and the linear impedance relation relation $v=p/\rho _{0}c_{0}$, the
equation for $\bar{T}$ is obtained by averaging of Eq.(\ref{heattransport})%
\begin{equation}
\frac{\partial \bar{T}}{\partial t}-\chi \nabla ^{2}\bar{T}=\frac{k^{2}b}{%
2\rho _{0}^{2}c_{0}^{2}}\left\vert P\right\vert ^{2},  \label{Teq}
\end{equation}%
where $\bar{T}$ represents the average temperature perturbation caused by
the wave energy dissipation. In Eq.(\ref{Teq}) the new parameter $b$ is
defined as $b=(\mu _{B}+4\mu /3)/\rho _{0}c_{p}$, and $\chi =\kappa /\rho
_{0}c_{p}$ is the thermal diffusivity.

On the other side, substituting Eqs.(\ref{quasiplane}) in Eqs.(\ref%
{continuityEq}) and (\ref{momentumtrans}) it follows that, in the parabolic
approximation, the pressure field is described by \cite{Naugolnykh} 
\begin{equation}
c_{0}\frac{\partial P}{\partial z}+\frac{\partial P}{\partial t}+\frac{%
ic_{0}^{2}}{2\omega }\nabla _{\bot }^{2}P+c_{0}aP=i\omega \frac{\gamma _{p}}{%
2}P\bar{T}  \label{eqP}
\end{equation}%
where $a=b\omega ^{2}/2\rho _{0}c_{0}^{3}$ and $\nabla _{\bot }^{2}$ is a
laplacian operator acting on transverse coordinates.

\subsection{Introducing the resonator}

\subsubsection{Modal equations}

In a resonator, the reflection at the boundaries imply the presence of
counterpropagating traveling waves,and the acoustic field $p$ can be
expressed in terms of forward and backward components 
\begin{equation}
P=P_{F}(\mathbf{r},t)e^{i(\omega t-kz)}+P_{B}(\mathbf{r},t)e^{i(\omega
t+kz)}.  \label{counter}
\end{equation}%
whose amplitudes are assumed to vary slowly in space and time (compared with
the evolution described by the exponential factors).

Substitution of expansion (\ref{counter}) into Eq.(\ref{Teq}), leads to the
following temperature distribution%
\begin{equation}
\bar{T}=T_{h}(\mathbf{r},t)+T_{g}(\mathbf{r},t)e^{i2kz}+T_{g}^{\ast }(%
\mathbf{r},t)e^{-i2kz},  \label{grating}
\end{equation}%
where $T_{0\text{ }}$is the homogeneous component and $T_{1}$ is the
amplitudes of the thermal grating induced by the acoustic wave. The acoustic
resonator containing a viscous fluid medium inside (e.g. glycerine), has
been previously addressed \cite{Lyakhov93}in the frame of plane-wave
approximation, we extend now the previous model by considering the sound
diffraction and temerature diffusion.

Projecting on the different longitudinal modes, the following equations for
the amplitudes are obtained: 
\begin{subequations}
\label{EquationsP}
\begin{eqnarray}
\frac{\partial P_{F}}{\partial t}+c_{0}\frac{\partial P_{F}}{\partial z}-i%
\frac{c_{0}^{2}}{2\omega }\nabla _{\bot }^{2}P_{F}+c_{0}aP_{F} &=&i\omega 
\frac{\gamma _{p}}{2}\left( T_{h}P_{F}+T_{g}^{\ast }P_{B}\right) , \\
\frac{\partial P_{B}}{\partial t}-c_{0}\frac{\partial P_{B}}{\partial z}-i%
\frac{c_{0}^{2}}{2\omega }\nabla _{\bot }^{2}P_{B}+c_{0}aP_{B} &=&i\omega 
\frac{\gamma _{p}}{2}\left( T_{h}P_{B}+T_{g}P_{F}\right) ,
\end{eqnarray}%
\end{subequations}
\begin{subequations}
\label{Temps}
\begin{eqnarray}
\frac{\partial T_{h}}{\partial t}-\chi \nabla _{\perp }^{2}T_{h} &=&\frac{%
k^{2}b}{2\rho _{0}^{2}c_{0}^{2}}\left( \left\vert P_{F}\right\vert
^{2}+\left\vert P_{B}\right\vert ^{2}\right) , \\
\frac{\partial T_{g}}{\partial t}+4\chi k^{2}T_{g}-\chi \nabla _{\perp
}^{2}T_{g} &=&\frac{k^{2}b}{2\rho _{0}^{2}c_{0}^{2}}P_{F}^{\ast }~P_{B},
\end{eqnarray}%
where $a=\omega ^{2}\left( \mu _{B}+4\mu /3\right) /(2\rho _{0}c_{0}^{3})$.

\subsubsection{Boundary Conditions.}

We consider an acoustic resonator with length $L$, filled\ with a viscous
medium, and bounded by reflecting surfaces, with intensity reflection
coefficients $\mathcal{R}$, located at $z=0$ and $z=L$. We assume that one
of the surfaces, vibrating at frequency $\omega $, acts as ultrasonic
source. In accordance with (\ref{counter}), boundary conditions satisfy 
\end{subequations}
\begin{eqnarray}
P_{F}(\mathbf{r}_{\bot },z &=&0;t)=\sqrt{\mathcal{R}}P_{B}(\mathbf{r}_{\bot
},z=0;t)+P_{in},  \label{cond1} \\
P_{B}(\mathbf{r}_{\bot },z &=&L,t)=\sqrt{\mathcal{R}}P_{F}(\mathbf{r}_{\bot
},z=L;t)e^{-i\delta },  \label{cond2}
\end{eqnarray}%
where the detuning parameter $\delta $ has been defined as%
\begin{equation}
\delta =2m\pi -2kL=2L\frac{\omega _{c}-\omega }{c_{0}},  \label{detuning}
\end{equation}%
where $m$ is an integer, selected in order to be $\omega _{c}$ the cavity
frequency that lies nearest to the driving frequency $\omega $.

\subsubsection{Mean Field Limit.}

When the reflectivity of the boundary walls is high, $\mathcal{R}\rightarrow
1$, the resulting model can be greatly symplified adopting the mean field
limit. Under this assumption, the two counterpropagating fields are
approximately constant and equal along the cavity (i.e. there is no spatial
modulation of the field envelopes), and one can assume that%
\begin{equation}
\bar{P}_{F}(\mathbf{r}_{\bot },t)=\bar{P}_{B}(\mathbf{r}_{\bot },t)\equiv P(%
\mathbf{r}_{\bot },t)  \label{meanfield}
\end{equation}%
where the overbar means the spatial average over the longitudinal coordinate 
$z$, 
\begin{equation}
~\bar{f}=\frac{1}{L}\int\limits_{0}^{L}dz~f(z).  \label{averageOp}
\end{equation}

We also consider that the frequency detuning between the driving and a
cavity mode is small, $2L\left( \omega _{c}-\omega \right) /c<<\pi $, which
allows to write the boundary condition Eq.(\ref{cond2}) as%
\begin{equation}
P_{B}(\mathbf{r}_{\bot },L,t)\approx \sqrt{\mathcal{R}}P_{F}(\mathbf{r}%
_{\bot },L,t)(1-i\delta ).
\end{equation}%
Averaging Eqs.(\ref{EquationsP}), with%
\begin{equation*}
\frac{1}{L}\int\limits_{0}^{L}\frac{\partial P_{f,b}}{\partial z}dz=\frac{%
P_{f,b}\left( L\right) -P_{f,b}\left( 0\right) }{L}.
\end{equation*}%
leads to%
\begin{eqnarray*}
\frac{\partial \bar{P}_{F}}{\partial t}+\frac{c_{0}}{L}\left[ P_{F}\left(
L\right) -P_{F}\left( 0\right) \right] -i\frac{c_{0}^{2}}{2\omega }\nabla
_{\bot }^{2}\bar{P}_{F}+c_{0}a\bar{P}_{F} &=&i\omega \frac{\gamma _{p}}{2}%
\left( T_{h}P_{F}+T_{g}^{\ast }P_{B}\right) , \\
\frac{\partial \bar{P}_{B}}{\partial t}-\frac{c_{0}}{L}\left[ P_{B}\left(
L\right) -P_{B}\left( 0\right) \right] -i\frac{c_{0}^{2}}{2\omega }\nabla
_{\bot }^{2}\bar{P}_{B}+c_{0}a\bar{P}_{B} &=&i\omega \frac{\gamma _{p}}{2}%
\left( T_{h}P_{B}+T_{g}P_{F}\right) ,
\end{eqnarray*}%
they can be reduced just to one equation%
\begin{equation}
\partial _{t}P=-\frac{c_{0}}{2\sqrt{\mathcal{T}}L}(\mathcal{T}+i\delta
)P_{T}+\frac{c_{0}}{2L}P_{in}+i\frac{c_{0}^{2}}{2\omega }\nabla _{\bot
}^{2}P-c_{0}aP+i\omega \frac{\gamma _{p}}{2}\left( \bar{T}_{h}+\frac{\bar{T}%
_{g}+\bar{T}_{g}^{\ast }}{2}\right) P,  \label{Pmean}
\end{equation}%
where the averaged temperature components as defined as $\bar{T}_{i}=%
\mathcal{L}T_{i}$ and we have used the property $\overline{\left( fg\right) }%
=\bar{f}~\bar{g}$ \cite{Bonifacio78}, valid in the mean field case. The
transmitted pressure can be assumed to be%
\begin{equation}
p_{T}(t)=\sqrt{\mathcal{T}}P(\mathbf{r}_{\bot };t)
\end{equation}%
and Eq.(\ref{Pmean}) is written as%
\begin{equation}
\partial _{t}P=-\Gamma (1+i\theta )P+P_{in}+i\frac{c_{0}^{2}}{2\omega }%
\nabla _{\bot }^{2}P+i\omega \frac{\gamma _{p}}{2}\left( \bar{T}_{h}+\frac{%
\bar{T}_{g}+\bar{T}_{g}^{\ast }}{2}\right) P,
\end{equation}%
where the parameters%
\begin{eqnarray}
\Gamma &\equiv &a~c_{0}+\mathcal{T}\frac{c_{0}}{2L}, \\
\theta &\equiv &\frac{\omega _{c}-\omega }{\Gamma }, \\
P_{in} &\equiv &\frac{c_{0}}{2L}p_{in},
\end{eqnarray}%
have been introduced. As well, under the same assumptions, Eqs.(\ref{Temps})
read 
\begin{subequations}
\label{Tmean}
\begin{align}
\frac{\partial \bar{T}_{h}}{\partial t}-\chi \nabla _{\perp }^{2}\bar{T}%
_{h}& =\frac{k^{2}b}{\rho _{0}^{2}c_{0}^{2}}\left\vert P\right\vert ^{2}; \\
\frac{\partial \bar{T}_{g}}{\partial t}+4k^{2}\chi \bar{T}_{g}-\chi \nabla
_{\perp }^{2}\bar{T}_{g}& =\frac{k^{2}b}{2\rho _{0}^{2}c_{0}^{2}}\left\vert
P\right\vert ^{2},
\end{align}

\subsection{The original model}

The mean-field model for the thermoacoustic resonator can be written as 
\end{subequations}
\begin{subequations}
\label{Model}
\begin{eqnarray}
\partial _{t}\bar{P} &=&-\frac{c_{0}}{2L}(\mathcal{T}+i\delta )\bar{P}+\frac{%
c_{0}}{2L}\bar{P}_{in}+i\frac{c_{0}^{2}}{2\omega }\nabla _{\bot }^{2}\bar{P}%
-c_{0}a\bar{P}+i\frac{\omega \gamma _{p}}{2}\left( \bar{T}_{h}+\bar{T}%
_{g}\right) \bar{P}, \\
\partial _{t}\bar{T}_{h} &=&-\gamma _{h}\bar{T}_{h}+\chi \nabla _{\perp }^{2}%
\bar{T}_{h}+\frac{k^{2}b}{\rho _{0}^{2}c_{0}^{2}}\left\vert \bar{P}%
\right\vert ^{2}, \\
\partial _{t}\bar{T}_{g} &=&-4k^{2}\chi \bar{T}_{g}+\chi \nabla _{\perp }^{2}%
\bar{T}_{g}+\frac{k^{2}b}{2\rho _{0}^{2}c_{0}^{2}}\left\vert \bar{P}%
\right\vert ^{2},
\end{eqnarray}%
where $\bar{T}_{g}$ has already been assumed to be real. We note that the
model is highly reminiscent of that for a coherently driven optical cavity
filled with a non-instantaneous (and nonlocal) Kerr medium \cite{Carmon04}.
Unlike its optical analog, in this thermoacoustic model there are two
nonlinearities (owed to $\bar{T}_{h}$ and to $\bar{T}_{g}$); notice however
that for $\gamma _{h}=4k^{2}\chi $ both nonlinearities behave as a single
one (a single temperature field $\bar{T}_{h}+\bar{T}_{g}$ can be defined so
that its evolution and that of the pressure just depend on themselves in
that case). Defining the following quantities 
\end{subequations}
\begin{eqnarray}
t_{\mathrm{p}}^{-1} &=&\frac{c_{0}\mathcal{T}}{2L}+c_{0}a,\;\Delta =t_{%
\mathrm{p}}\left( \omega _{c}-\omega \right) ,\;\tau _{\mathrm{p}}=\gamma _{%
\mathrm{h}}t_{\mathrm{p}},\;\tau _{\mathrm{g}}=\frac{\gamma _{\mathrm{h}}}{%
4k^{2}\chi } \\
\tau &=&\gamma _{\mathrm{h}}t,\;\nabla ^{2}=\frac{t_{\mathrm{p}}c_{0}^{2}}{%
2\omega }\nabla _{\bot }^{2},\;D=\frac{\chi }{\gamma _{\mathrm{h}}t_{\mathrm{%
p}}}\frac{2\omega }{c_{0}^{2}},\;P_{\mathrm{in}}=t_{\mathrm{p}}\frac{c_{0}}{%
2L}\sqrt{\frac{\gamma _{p}\omega \tau _{\mathrm{p}}k^{2}b}{4\gamma _{\mathrm{%
h}}\rho _{0}^{2}c_{0}^{2}}}\bar{P}_{in} \\
P &=&\sqrt{\frac{\gamma _{p}\omega \tau _{\mathrm{p}}k^{2}b}{4\gamma _{%
\mathrm{h}}\rho _{0}^{2}c_{0}^{2}}}\bar{P},\;H=\frac{\gamma _{p}\omega \tau
_{\mathrm{p}}}{2}\bar{T}_{\mathrm{h}},\;G=\frac{\gamma _{p}\omega \tau _{%
\mathrm{p}}}{2}\bar{T}_{\mathrm{g}},
\end{eqnarray}%
the model equations, Eqs.(\ref{Model}) become 
\begin{subequations}
\label{ThAcModel}
\begin{eqnarray}
\tau _{\mathrm{p}}\partial _{\tau }P &=&-\left( 1+i\Delta \right) P+P_{%
\mathrm{in}}+i\nabla ^{2}P+i\left( H+G\right) P,  \label{dP} \\
\partial _{\tau }H &=&-H+D\nabla ^{2}H+2\left\vert P\right\vert ^{2},
\label{dH} \\
\partial _{\tau }G &=&-\tau _{\mathrm{g}}^{-1}G+D\nabla ^{2}G+\left\vert
P\right\vert ^{2}.  \label{dG}
\end{eqnarray}%
We consider a resonator with high quality plates ($\mathcal{T}=0.1$),
separated by $L=5$ cm, driven at a frequency $f=2$MHz and containing
glycerine at 10%
${{}^o}$%
C. Under these conditions the medium parameters are $c_{0}=2\times 10^{3}$m s%
$^{-1},\alpha _{0}=10$ m$^{-1},\rho _{0}=1.2\times 10^{3}$kg m$%
^{-3},c_{p}=4\times 10^{3}$J kg$^{-1}$K$^{-1}$, $\sigma =10^{-2}$ K$^{-1}$
and $\kappa =0.5$ W m$^{-1}$K$^{-1}$($\chi =10^{-7}$m$^{2}$s$^{-1})$. In
this case $t_{p}=2\times 10^{-5}$ s, $t_{g}=6\times 10^{-2}$ s, and our
length unit is $l_{d}=2$ mm. For a resonator with a large Fresnel number,
the relaxation of the homogeneous the temperature is mainly due to the heat
flux through the boundaries, and can be estimated from the Newton's colling
law as $t_{h}\sim 10^{1}$ s. Then, under usual conditions we get the
normalized decay times $\tau _{p}\sim 10^{-6}$, and $\tau _{g}\sim 10^{-2}$,
and the diffusion constant $D\sim 10^{0}$.Note that for $\tau _{\mathrm{g}%
}=1 $ a single temperature field ($H+G$) can be defined, whose evolution
depends just on $P$ and vice versa. This model exhibits a rich
spatiotemporal dynamics and exhibits pattern formation and localized
structure formed in the transverse cross-section of the cavity. Some of
these results are studied in \cite{arxiv}.

\end{subequations}

\end{document}